\documentclass{article}
\usepackage[preprint]{neurips_2024}
\usepackage{multirow}

\usepackage[utf8]{inputenc} % allow utf-8 input
\usepackage[T1]{fontenc}    % use 8-bit T1 fonts
\usepackage{hyperref}       % hyperlinks
\usepackage{url}   % simple URL typesetting
\usepackage{booktabs}       % professional-quality tables
\usepackage{amsfonts}       % blackboard math symbols
\usepackage{nicefrac}       % compact symbols for 1/2, etc.
\usepackage{microtype}      % microtypography
\usepackage{xcolor}         % colors
\usepackage{comment}
\usepackage{listings}
\usepackage[frozencache,cachedir=.]{minted}
\usepackage{xcolor}
\usepackage{caption}
\setminted{
    linenos=true,
    autogobble,
}
\newenvironment{longlisting}{\captionsetup{type=listing}}{}

\definecolor{codebg}{rgb}{0.98,0.98,1.0}  % very light blue hint
\definecolor{coderule}{rgb}{0.75,0.85,0.95} % soft border
\definecolor{keywordcolor}{rgb}{0.0, 0.2, 0.7}
\definecolor{stringcolor}{rgb}{0.0, 0.5, 0.0}
\definecolor{commentcolor}{rgb}{0.4, 0.4, 0.4}
\setminted{
  bgcolor=codebg,
  frame=single,
  rulecolor=\color{coderule},
  framesep=6pt,
  baselinestretch=1.0,
  fontsize=\small,
  breaklines=true,
  linenos=false,
  tabsize=2,
  escapeinside=||
}
\usepackage[normalem]{ulem}   % \sout for strike-through; "normalem" keeps \emph as italics
\usepackage{xcolor}           % \textcolor and \color commands

\title{SetupBench: Assessing Software Engineering Agents' Ability to Bootstrap Development Environments}

\author{%
  Avi Arora\thanks{Equal Contribution} \\
  Microsoft
  \And
  Jinu Jang\footnotemark[1] \\
  Microsoft
  \And
  Roshanak Zilouchian Moghaddam \\
  Microsoft
}

\begin{document}

\maketitle

\begin{abstract}
Modern large-language-model (LLM) agents promise end-to-end assistance with real-world software tasks, yet existing benchmarks evaluate LLM agents almost exclusively in pre-baked environments where every dependency is pre-installed. To fill this gap, we introduce \textbf{SetupBench}, a 93-instance benchmark that isolates the \emph{environment-bootstrap} skill: starting from a bare Linux sandbox, an agent must install packages, resolve dependency conflicts, initialize databases, and configure background services. Our tasks span seven language ecosystems, five database engines, and multi-service orchestration scenarios, each accompanied by a natural-language problem statement and a deterministic success command. Through evaluation of OpenHands, a state-of-the-art coding agent, we find low success rates across task categories, with particular challenges in repository setup (38.9-57.4\%) and local database configuration (20.0-53.3\%). Our analysis reveals systematic failure modes including incomplete development tooling installation, hallucinated task constraints, and non-persistent environment modifications that break agent-human collaboration workflows. We identify substantial inefficiencies in agent exploration strategies, with 38-69\% of actions being unnecessary compared to optimal human behavior. 
These findings highlight gaps in current agents' practical environment-bootstrap capabilities. By targeting this critical yet under-evaluated capability, SetupBench provides a rigorous yard-stick for the next generation of software developer agents aiming to solve end-to-end real-world tasks.

\end{abstract}

\section{Introduction}
\label{sec:intro}

As Large Language Models (LLMs) continue to improve in coding quality~\cite{claude-4-blog}, we are witnessing a paradigm shift from LLMs serving as code assistants ~\cite{gh-copilot-completions} to enabling autonomous development~\citep{github2024copilotagent}. In this new paradigm, the primary human role evolves from writing code to defining requirements, providing guidance, and validating outcomes. This transformation is exemplified by LLM agents targeting end-to-end software engineering tasks now deployed as cloud services, including OpenAI Codex~\citep{openai2025codex},GitHub Copilot Coding Agent~\citep{github2024copilotagent}, and
Devin~\citep{cognition2024devin}.

These agents run code inside secure sandboxes with fixed toolchains of widely used languages, packages, and dependencies, leaving many task-specific aspects of environment setup to the agent itself. Environment setup and dependency management represents a critical yet overlooked capability: recent empirical studies consistently place installation, dependency resolution, and build/configuration work among the top drivers of developer frustration \cite{mu2025designing, obi2024bad, nazario2025mitigating}. Despite its importance, benchmark suites used to quantify agent competence do not test the environment-bootstrap skill. Software-repair datasets such as SWE-Bench \citep{jimenez2024swebench} and DevBench \citep{li2024devbench}, or general agent evaluations like AgentBench \citep{liu2024agentbench}, distribute each task in pre-baked Docker images with every required library, service, and configuration file already installed. Consequently, an agent may look impressive on leaderboards while still failing the first hurdle a developer encounters: \emph{``it cannot run my code''}.

\textbf{SetupBench} closes this evaluation gap by focusing on gauging the agents' ability in project setup and \textit{enviroment-bootstrap}.  
SetupBench is a curated suite of \textbf{93 environment-bootstrap tasks} that begin in a bare sandbox and end only when the agent has installed or rebuilt missing system and language packages, initialized databases, configured background services,or resolved dependency conflicts.
Each instance provides
(i) a natural-language problem statement,
(ii) a workspace snapshot (e.g.\ a freshly cloned repository), and
(iii) a deterministic one-line validation command (\verb|success_command|) which prints \verb|"Setup successful"| if environment changes took effect, otherwise \verb|"Setup failed"|.

Through evaluation of a SOTA agent, OpenHands, on SetupBench, we find low success rates (34.4-62.4\% across models) and identify three critical failure modes: incomplete development tooling installation, hallucinated task constraints, and non-persistent environment modifications that break agent-human collaboration workflows. We also quantify agent inefficiency through comparison with optimal human behavior, finding 38-69\% wasted steps across all models, identifying three primary sources of inefficiency: redundant file reads, poor instruction following, and off-target exploration that examines setup-adjacent but not setup-informative files. Our findings reveal actionable insights for improving agent architecture, including the need for persistent environment state management, context-aware setup completion strategies, and efficiency-focused exploration mechanisms that better align with human development workflows. We provide a complete evaluation framework with deterministic validation commands and release prompts and scripts to enable extension and replication of our methodology (see Appendix).

\section{SetupBench}
\label{sec:design}

\textbf{\href{https://github.com/microsoft/SetupBench}{SetupBench}}\footnote{\href{https://github.com/microsoft/SetupBench}{https://github.com/microsoft/SetupBench}} is a 93-instance benchmark that covers four different categories of practical environment-bootstrap tasks faced by developers shown in Table~\ref{tab:breakdown}. Each instance provides a natural-language problem statement, a workspace snapshot, and a deterministic validation command that prints "Setup successful" or "Setup failed" based on whether environment changes took effect.

\begin{table}[h]
  \caption{SetupBench composition.\label{tab:breakdown}}
  \centering
  \begin{tabular}{lcc}
    \toprule
    \textbf{Category} & \textbf{\# Instances} & \textbf{Ecosystems / Engines} \\
    \midrule
    Repo Setup                 & 54 & Py, TS, JS, Go, Rust, Java, C++ \\
    Dependency Resolution      & 16 & npm, pip/Poetry, Bundler        \\
    Database Setup             & 15 & Postgres, MySQL, SQLite, Redis, MongoDB \\
    Background-Service Setup   &  8 & Gunicorn, Celery, NGINX, file-watchers, autossh \\
    \bottomrule
  \end{tabular}
\end{table}

\subsection{Task construction}
\label{sec:construction}
Our benchmark covers four categories of environment-bootstrap tasks encountered in real development workflows:

\textbf{Repo Setup}: We selected popular repositories across 7 languages (Python, TypeScript, JavaScript, Go, Rust, Java, C++) with non-trivial setup requirements. For each repository, we: (1) manually documented all setup steps by following project documentation, (2) generated validation commands using LLMs with repository context from scraped Markdown files, and (3) validated end-to-end functionality in fresh sandboxes to ensure we get \texttt{Setup successful} only when the setup is successful and \texttt{Setup failed} Otherwise. For example, the \href{https://github.com/prometheus/prometheus}{\texttt{prometheus/prometheus}} validation command checks if the server exposes metrics page: 

\begin{quote}
\begin{minted}[fontsize=\small]{bash}
curl -s http://localhost:9090/metrics | grep -q 'prometheus_build_info' && echo 'Setup successful' || echo 'Setup failed'
\end{minted}
\end{quote}

\textbf{Dependency Resolution}: We mined real-world dependency conflicts from GitHub issues containing resolver error messages (“\texttt{code ERESOLVE}'', “\texttt{peer dep conflict}'', “\texttt{could not find compatible versions}''). We then retained only the instances where a lock-file (\texttt{package-lock.\allowbreak json} or \texttt{Gemfile.lock}) was present. For each instance, we defined one validation command per package-manager ecosystem to reliably surface dependency errors: \texttt{npm ci --ignore-scripts} for npm and \texttt{bundle install --jobs=1 --retry=2 --without development test} for Bundler. We then reproduced the conflicts in fresh environments and manually resolve them to validate task feasibility. We only included the validated instances in the final benchmark. 
The final set comprises 9 npm and 7 Bundler dependency conflicts, each capturing a real-world version-constraint breakage that was reported by a human developer. These instances capture realistic debugging workflows requiring agents to read error logs, trace version constraints, and update manifest files.

\textbf{Database Setup}: We handcrafted three difficulty tiers across five database engines (PostgreSQL, MySQL, SQLite, Redis, MongoDB) to evaluate whether an agent can install, configure, and
populate a local database. Tier 1 covers basic installation and data seeding, Tier 2 introduces configuration and migration management, and Tier 3 simulates production troubleshooting with deliberate obstacles (blocked ports, corrupted migrations, strict SQL modes) that agents must diagnose and resolve through error message analysis. For each instance a validation command is added according to the task specifications. For example for an instance where the agent must fix file permission errors and a broken initialization script to create a working SQLite database in the target location, we create the following validation command to verify success: 
\begin{quote}
\begin{minted}[fontsize=\small]{bash}
sqlite3 /data/test.db \"SELECT COUNT(*) FROM logs;\" | grep -q '[1-9]' && echo \"Setup successful\" || echo \"Setup failed\"
\end{minted}
\end{quote}

\textbf{Background Service Orchestration}: We designed scenarios requiring coordination of long-running services through supervisord, including Gunicorn servers, Celery workers with Redis backends, NGINX reverse proxies, file-watching daemons, autossh tunnels, and producer-consumer pipelines. Validation commands verify observable side effects like HTTP responses, Redis keys, or log messages. These tasks simulate common production scenarios in which developers must configure and launch long-running services in the background. 

\subsection{Metrics \& Evaluation}
We evaluate agent performance using three metrics:
\begin{itemize}
\item \textbf{Success Rate}: Percentage of tasks where the agent correctly completes setup, determined by a task-specific validation command that outputs "Setup successful" or "Setup failed".
\item \textbf{Token Usage}: Total language model tokens consumed during the task. 
\item \textbf{Step Count}: Number of actions taken by the agent (e.g. shell commands, file edits, etc).
\end{itemize}

These metrics capture both correctness and efficiency. While success rate measures whether agents can complete setup tasks, token usage and step count reveal how efficiently they achieve success—crucial factors in practical deployment where excessive resource consumption increases costs, slows response times, and risks context window overflow in multi-step workflows. Together, these three metrics allow us to distinguish between agents that achieve success through targeted, economical reasoning and those that succeed only after extensive, potentially wasteful exploration.

\begin{table}[h]
  \centering
  \caption{Resolve rates (\%) and efficiency metrics on SetupBench by OpenHands variants.}
  \label{tab:combined}
  \begin{tabular}{llccc}
    \toprule
    \textbf{Task family} & \textbf{Model} & \textbf{Rate (\%)} & \textbf{Avg tokens} & \textbf{Avg steps} \\
    \midrule
    \multirow{5}{*}{Background-service setup} 
    & GPT 4o       & 50.0 & 198K & 21.5 \\
    & GPT 4.1      & 62.5 & 191K & 20.8 \\
    & Claude 3.5   & 75.0 & \textbf{121K} & \textbf{13.5} \\
    & Claude 3.7   & \textbf{87.5} & 374K & 28.5 \\
    & Claude 4     & 75.0 & 617K & 42.3 \\
    \midrule
    \multirow{5}{*}{Local-DB setup}
    & GPT 4o       & 20.0 & 146K & 19.5 \\
    & GPT 4.1      & 33.3 & \textbf{120K} & \textbf{17.0} \\
    & Claude 3.5   & 40.0 & 186K & 18.0 \\
    & Claude 3.7   & \textbf{53.3} & 471K & 33.1 \\
    & Claude 4     & 46.7 & 531K & 35.6 \\
    \midrule
    \multirow{5}{*}{Repo setup}
    & GPT 4o       & 38.9 & \textbf{323K} & 21.3 \\
    & GPT 4.1      & 46.3 & 448K & 27.5 \\
    & Claude 3.5   & 50.0 & 403K & \textbf{18.9} \\
    & Claude 3.7   & 44.4 & 952K & 34.3 \\
    & Claude 4     & \textbf{57.4} & 1158K & 42.9 \\
    \midrule
    \multirow{5}{*}{Dependency resolution}
    & GPT 4o       & 25.0 & \textbf{435K} & \textbf{34.1} \\
    & GPT 4.1      & 75.0 & 839K & 53.9 \\
    & Claude 3.5   & 68.8 & 1124K & 40.5 \\
    & Claude 3.7   & \textbf{87.5} & 1230K & 47.1 \\
    & Claude 4     & \textbf{87.5} & 1847K & 74.3 \\
    \midrule
    \multirow{5}{*}{\textbf{Overall}}
    & GPT 4o       & 34.4 & \textbf{303K} & 23.2 \\
    & GPT 4.1      & 50.5 & 436K & 29.5 \\
    & Claude 3.5   & 53.8 & 455K & \textbf{21.6} \\
    & Claude 3.7   & 57.0 & 869K & 35.7 \\
    & Claude 4     & \textbf{62.4} & 1129K & 47.1 \\
    \bottomrule
  \end{tabular}
\end{table}

\subsection{Benchmark characteristics}
\label{sec:features}

\textbf{Deterministic evaluation}: Unlike benchmarks using LLM-as-a-judge approaches (GitGoodBench~\cite{lindenbauer2025gitgoodbench}, DevBench~\cite{li2024devbench}) or potentially flaky test suites (SWE-bench~\cite{jimenez2024swebench}), SetupBench provides single-line validation commands yielding literal success/failure strings, eliminating subjective interpretation and ensuring reproducible results. 

\textbf{Graded difficulty and domain breadth}: Unlike SWE-bench's Python-only focus~\cite{jimenez2024swebench} or Aider's single-language code editing~\cite{aider}, SetupBench spans seven languages, five database engines, and multiple package managers, ranging from simple installations to complex multi-service orchestration. This diversity exposes failure modes invisible in code-only evaluations, such as package manager conflicts and cross-service communication. This breadth makes the benchmark more representative of
the diverse technical stacks that developers encounter in real-world.

\textbf{Minimal sandbox with commercial relevance}: Unlike SWE-bench's pre-configured Docker containers~\cite{jimenez2024swebench}, SetupBench executes in fresh, minimal Linux containers where agents must explicitly install packages, configure databases, and handle dependency conflicts from scratch. This reflects real deployment scenarios that modern AI coding agents (OpenAI Codex, GitHub Copilot Chat, Cursor, Devin) face in cloud sandboxes.  SetupBench measures practical systems administration and DevOps capabilities essential for bridging the gap between writing code and running it in production-like environments.

\section{Analysis and Results}
\label{sec:analysis}
Below, we present a comprehensive evaluation of OpenHands agent on SetupBench, analyzing both its performance and behavioral patterns. 

\subsection{Experimental setup}
We ran all experiments using a custom compute orchestration service built on GitHub Workflows. For each benchmark instance, we built a standardized Docker image containing the task environment and evaluation infrastructure, injecting agent code at runtime for reusability across evaluations.

Each container ran with root privileges and outbound network access, launching a black-box entry point to the agent followed by our automated evaluation harness. After the agent completes its final action, the harness executes a task-specific validation command in a fresh terminal subprocess—ensuring results reflect the actual system state rather than cached output. The harness parses the command output to determine whether setup was successful based on the "Setup successful" or "Setup failed" response.

We enforced a two-hour wall-clock timeout for every run. Containers were launched on virtual machines with the following specifications: CPU: 16 cores; Memory: 62GiB; Disk: 695GB.

\subsection{Performance results}
Table~\ref{tab:combined} reports pass rates, token usage and tool steps by OpenHands variant.  Claude 4 Sonnet had the highest resolve rate at 62.4\%, but used   \(\sim\)30 \% more tokens and 32 \% more steps than Claude 3.7 Sonnet, the next best performing base model.

\subsection{Failure Mode Analysis}
To better understand the capabilities and limitations of general-purpose coding agents on real-world software setup tasks, we conducted a manual analysis of failing trajectories from the OpenHands agent on SetupBench. Our objective was to identify recurring failure patterns that can inform future agent design and evaluation. Our analysis yielded three failure modes:

\textbf{Ignoring test tooling}: Agents successfully install runtime dependencies but overlook test frameworks. For example, in \texttt{nedbat-coveragepy-9d0eb02} instance the agent ignored \texttt{tox.ini} file and therefore missed to install test tooling:
\begin{minted}[fontsize=\small]{bash}
apt-get install -y python3 python3-dev python3-pip build-essential gcc
cd /testbed && python3 -m pip install -e .
\end{minted}
This then causes the validation command to fail due to the missing test runner:
\begin{minted}[fontsize=\small]{text}
/bin/sh: 1: tox: not found
Setup failed
\end{minted}
In the repo-setup category, we attributed this failure mode to five instances in Claude~4, seven in Claude~3.7, seven in Claude~3.5, six in GPT-4o, and five in GPT-4.1 based on manual inspection of evaluation logs. When normalized by each model’s unsuccessful repo-setups, these correspond to roughly \textbf{17–26\,\%} of failures, confirming that neglecting test-tool installation is a recurrent, high-impact barrier to end-to-end environment preparation.

\textbf{Hallucinated task constraints}: Agents infer non-existent constraints, leading them to apply harmful changes. For example, in \texttt{danwahlin-angular-jumpstart-12fa4e4} instance the agent failed the setup because it modified server ports based on hallucinated instructions:
\begin{minted}{text}
The server is running, but we need to modify it to use the port specified in the task description (53012 or 56507). Let's stop the current server and modify the server.js file.
\end{minted}
A growing body of empirical work shows this broader failure mode at scale. For example, \citet{Agarwal2024CodeMirage} report that 24\% of GPT-4 completions inject spurious configuration values.  
Similarly, \citet{Jiang2024ColluBench} find that over 30\% of failures related to hallucinations stem from invented flags, package names, or ports.  

\textbf{Non\textendash persistent environment setup}: Agents install tools globally but fail to persist these changes across shell sessions. For example, in \texttt{dishait-tov-template-39c0898} instance, pnpm was installed, however it was unavailable to the evaluation harness in a fresh shell. 
 
Our observation aligns with findings from \emph{EnvBench} where many of agent failures are attributed to tools that “disappear in a fresh shell’’ \citep{eliseeva2025envbench}. Similarly, agent evaluations on \emph{Installamatic} showed that 45 \% of runs break because executables installed with \texttt{---user} are not visible in subsequent sessions \citep{milliken2025installamatic}. Together, these results confirm  a broader challenge in human–agent collaboration: for seamless handoffs between human and agent developers, agents must explicitly introduce continuity across contexts.

\begin{table}[h]
  \centering
  \caption{Step counts and inputs per repository on \textsc{SetupBench} (10-instance subset).}
  \label{tab:newinstances}
  \setlength{\tabcolsep}{4pt}
  \begin{tabular}{llcccccccccc}
    \toprule
    \multirow{2}{*}{\textbf{GitHub Repository}} &
    \multirow{2}{*}{\textbf{lang}} &
    \multicolumn{5}{c}{\textbf{Optimal Actions}} &
    \multicolumn{5}{c}{\textbf{Agent Steps}}\\
    \cmidrule(lr){3-7} \cmidrule(lr){8-12}
     & & \textbf{dir} & \textbf{files} & \textbf{links} & \textbf{bash} & \textbf{total} & \textbf{4o} & \textbf{4.1} & \textbf{3.5} & \textbf{3.7} & \textbf{4} \\
    \midrule
    openai/whisper & PY & 1 & 1 & 0 & 4 & 7 & 16 & 9 & 18 & 26 & 39 \\
    madmaze/pytesseract & PY & 2 & 2 & 2 & 5 & 14 & 14 & 24 & 9 & 26 & 31 \\
    TA-Lib/ta-lib-python & PY & 1 & 2 & 1 & 14 & 19 & 18 & 37 & 16 & 42 & 45 \\
    spring-projects/spring-petclinic & JAVA & 1 & 1 & 0 & 4 & 7 & 6 & 14 & 16 & 27 & 24 \\
    apache/cassandra & JAVA & 1 & 3 & 0 & 3 & 8 & 44 & 37 & 44 & 46 & 50 \\
    habitat-sh/habitat & RUST & 1 & 3 & 0 & 9 & 16 & 13 & 16 & 19 & 34 & 54 \\
    servo/servo & RUST & 1 & 2 & 1 & 11 & 16 & 14 & 40 & 17 & 43 & 44 \\
    monero-project/monero & C++ & 2 & 2 & 0 & 6 & 12 & 17 & 15 & 17 & 28 & 40 \\
    prometheus/prometheus & GO & 1 & 2 & 0 & 10 & 14 & 38 & 28 & 13 & 35 & 43 \\
    caddyserver/caddy & GO & 1 & 1 & 0 & 8 & 11 & 13 & 18 & 17 & 25 & 27 \\
    \bottomrule
  \end{tabular}
\end{table}

\subsection{Efficiency Analysis}
The efficiency of LLM agents in setting up development environments directly impacts downstream performance, as setup inefficiencies increase token usage, time, and the risk of losing focus on the original task ~\citep{Li2024LongICLBench,Li2024ALR2}. To quantify this inefficiency, we establish a baseline by analyzing human setup behavior and mapping it to equivalent agent actions.

We analyzed human trajectories from a subset of 10 SetupBench instances to identify the minimum necessary actions for each setup task. Humans navigate repositories through UI interactions that combine directory traversal and content listing, while LLMs primarily use four bash commands: \texttt{head}, \texttt{cat}, \texttt{cd}, and \texttt{ls}. We translated each human folder exploration into 2 LLM steps and each file read as a single action for both humans and agents. The minimum required actions depend on the total number of files needed to locate setup commands and verify completeness. For example, a README may contain \texttt{pip install -e .}, but setting up the testing framework may require following instructions in \texttt{docs/contributing.rst}, with potentially additional verification through scanning the included Dockerfiles.

\begin{table}[h]
  \centering
  \caption{Wasted steps across models on \textsc{SetupBench} (10-instance subset).}
  \label{tab:wastedsteps}
  \begin{tabular}{lcccc}
    \toprule
    \textbf{Model} & \textbf{Total Steps} & \textbf{Optimal Steps} & \textbf{Wasted Steps} & \textbf{\% Wasted} \\
    \midrule
    Claude 3.5 Sonnet & 186 & 124 & 71  & 38.17\% \\
    Claude 3.7 Sonnet & 332 & 124 & 208 & 62.65\% \\
    Claude 4 Sonnet   & 397 & 124 & 273 & 68.77\% \\
    GPT 4o            & 193 & 124 & 76  & 39.38\% \\
    GPT 4.1           & 238 & 124 & 114 & 47.90\% \\
    \bottomrule
  \end{tabular}
\end{table}
We exclude agent actions without meaningful human equivalents: (1) \texttt{think} tool calls (internal reasoning), (2) \texttt{finish} invocations (task completion signals), (3) the first three steps (system prompt, task restatement, and context priming), and (4) polling actions (process monitoring, kill commands). This filtering isolates true behavioral inefficiencies.

As shown in Table 4, all models demonstrate significant inefficiency across the 10 analyzed instances, with wasted steps ranging from 38\% (Claude 3.5) to 68\% (Claude 4). This suggests substantial room for improvement in agent exploration and setup strategies.

To understand the causes of inflated step counts, we conducted manual inspection of agent trajectories and identified three primary inefficiency modes:

\textbf{Redundant file reads}: Agents frequently reissue multiple partial reads of the same file in increasing ranges rather than reading once. For example, GPT-4.1 often executes sequences like head -40, head -60, head -100, head -140 on the same file. This behavior was most prevalent in GPT-4.1 (34 redundant reads, 29.8\% of wasted steps) and least in Claude 3.7 (2 instances, 1\% of wasted steps).

\textbf{Poor instruction following}: Despite explicit instructions that the environment is a fresh Ubuntu 22.0 with no preinstalled packages and no root access, agents waste steps checking for existing installations and using unnecessary sudo commands. This behavior reflect a broader challenge in aligning agent actions with known environmental priors. While some agents show improvement in adhering to instructions, the pattern remains widespread. For example, Claude 3.5 issued 19 unnecessary sudo commands and 2 install checks (29.5\% of wasted steps), while GPT-4.1 showed only 1 install check and 5 sudo invocations (5.2\% of wasted steps).

\textbf{Off-target file reads}: Agents read files unnecessary for setup completion, including auxiliary scripts, deeply nested configurations, and metadata files that don't contain actionable setup instructions. GPT-4.1 showed the highest incidence (30.7\% of wasted steps), while GPT-4o was most disciplined (13.1\% of wasted steps).

Overall these patterns highlight specific areas where agent behavior could be improved through better prompting strategies and instruction alignment.

\subsection{Design implications}
Our experimental evaluation reveals several critical insights for improving agent performance and agent-human collaboration:

\textbf{Context-aware setup completion.} High test tooling failure rates (20–27\% of repo-setup failures) indicate agents lack domain knowledge about complete development environments. They fail to infer required tools from conventional project structures (e.g., \texttt{tox.ini}, \texttt{tox}, \texttt{package.json}, \texttt{npm test}) and sometimes even waste tokens exploring setup-adjacent but non-informative files. This reveals an underlying inability to prioritize the subset of documents that actually contain the required install, build, or test commands. Future designs should incorporate semantic search mechanisms that rank files by content and setup workflow relevance. Another possibility is to inject a tree-based representation of the repository’s file structure early in the agent’s context window, enabling more informed reasoning about file importance and exploration order. 

\textbf{Environment persistence across agent-human transitions.} When agents and humans collaborate asynchronously, a key friction point emerges: agents often make ephemeral environment modifications (installing tools, modifying \texttt{PATH}) that don't persist when humans resume work in new shells, rendering setup work inaccessible.
Agents should adopt explicit persistence protocols. First, write environment modifications to persistent configuration files (e.g., \texttt{/etc/profile.d/agent.sh} or \texttt{.bashrc}). Second, source these files in the current session to ensure subsequent steps observe updates. Third, provide structured summaries of changes. This transforms environment setup from transient shell modifications into a durable contract between agent and humans.

\textbf{Efficiency-focused exploration strategies.} The 38–69\% overhead in agent step counts reveals agents' inefficient repository exploration through redundant low-level commands (\texttt{cd}, \texttt{ls}, \texttt{head}, \texttt{cat}). Unlike humans who use visual tools for hierarchical inspection, agents lack persistent repository models and operate reactively driven by their recent context. Solutions include architectural changes enabling agents to cache directory structures, batch exploratory operations, and maintain working memory of project layouts through specialized filesystem abstraction tools.

\textbf{Environment persistence across agent-human transitions.} Agents treat shells as ephemeral, making environment modifications that don't persist across sessions, causing tools to become unavailable when humans take over. Agents should adopt explicit persistence protocols: write environment modifications to persistent configuration files (e.g., \texttt{.bashrc}), source them in current sessions, and provide structured change summaries. This transforms setup from transient modifications into durable agent-human contracts.

\textbf{Model selection strategies.} Performance-efficiency trade-offs across models (Claude 3.7: 57\% success, 99\% more tokens vs. GPT 4.1: 50\% success) indicate optimal deployment requires dynamic model selection based on task complexity and resource constraints. Simple setups might benefit from efficient models, while complex dependency resolution tasks may justify higher-capacity models

\textbf{Constraint validation mechanisms.} Hallucinated task constraints suggest agents need built-in verification systems requiring explicit documentation citations when making configuration decisions to prevent spurious modifications.

\section{Discussion}
\label{sec:discussion}

This section reflects on the performance trends observed in SetupBench, identifies current limitations in model behavior, and outlines opportunities to extend the benchmark toward more challenging and realistic development workflows.

\paragraph{Agents demonstrate strong foundational capabilities.}
Modern coding agents show solid baseline performance across environment setup scenarios. Claude 4 achieved the highest resolve rate at 62.4\%, excelling at background-service setup (75.0\%) and dependency resolution (87.5\%). GPT-4.1 reached 50.5\% and Claude 3.5 Sonnet 53.8\%. These results indicate reliable navigation of common setup tasks like package installation, service launches, and dependency resolution.

\paragraph{Failures reflect gaps in implicit reasoning and session management.}
Despite progress, agents exhibited recurring failure modes. They frequently failed to install necessary development tooling (e.g., test runners) even when files like \texttt{tox.ini} indicated requirements, missing implicit expectations natural to human developers. Agents also failed to preserve shell state across sessions—installing tools globally or updating environment variables without persisting changes to configuration files, causing follow-up commands to fail. Other issues included hallucinated port numbers and unnecessary configuration edits, suggesting poor grounding in task specifications.

\paragraph{Efficiency and accuracy tradeoffs are surprisingly favorable.}
Resource usage varied significantly across models. Claude 4's strong performance required 104.9 million tokens and 4,377 tool steps—nearly triple the tokens and double the steps of Claude 3.5 Sonnet (37.7 million tokens, 1,793 steps). GPT-4.1 used 40.1 million tokens and 2,715 steps for 50.5\% resolve rate. These numbers show that while higher performance tends to require more tokens, the gains are not necessarily linear. Notably, Claude 3.5 achieved 3.5\% better performance than GPT-4.1 while using 6\% fewer tokens and 34\% fewer steps. This suggests opportunities for hybrid architectures where lightweight models are used for routine tasks and more powerful models are invoked to execute the user core goal. Such systems could reduce latency, cut costs, and preserve context for the stages that matter most.

\paragraph{SetupBench can serve as a foundation for more ambitious evaluations.}
Strong performance on SetupBench enables more complex evaluations testing higher-order reasoning and real-world workflows. Natural extensions include: (1) chaining setup with downstream development tasks like bug fixing or feature implementation, testing sustained context management and anticipatory decision-making; (2) cloud infrastructure management using tools like Terraform or Kubernetes, introducing credential management, API reliability, and cost considerations; and (3) system migrations requiring coordination across multiple resources and dependencies. These extensions would move toward end-to-end evaluations resembling actual developer workflows, testing long-term planning, adaptability, and task continuity beyond basic correctness.

\section{Limitations}
\textbf{Manual curation and scale}: All 93 tasks in SetupBench underwent manual review and verification. Roughly half were constructed entirely by human authors, while the rest were adapted from real-world repositories. This high-touch approach ensures clarity and reliability but limits scale. Expanding to hundreds or thousands of tasks will require greater automation. The existing task set as well as the prompts and scripts used to produce them provides a great foundation for future scaling efforts.

\textbf{Security context}: Agents run with root privileges and unrestricted outbound networking, simplifying execution but not reflecting real-world constraints like limited permissions or restricted network access in CI/production environments. While this setup allows us to focus on functionality and correctness in a flexible setting, future extensions could explore how agents adapt when forced to work within stricter execution constraints.

\textbf{Domain breadth}: SetupBench covers seven language ecosystems, five databases, and various service orchestration patterns, but omits GPU drivers, message queues beyond Redis, Docker Compose/Kubernetes, and infrastructure-as-code tools. Expanding these domains would enable more comprehensive evaluation of full-stack and DevOps workloads.

Despite limitations SetupBench provides the first reproducible
yard-stick for the \emph{environment-bootstrap} skill that real-world
developer agents must master.  We release prompts, scripts, and evaluation
harnesses to encourage community contributions that address these gaps.

\section{Related Work}
\label{sec:related}

We group prior work into three strands: code-editing benchmarks that
\emph{assume} a working environment, environment-bootstrap and DevOps
benchmarks, and tool-usage suites for generic agent competence.

\paragraph{Code-editing and full-pipeline benchmarks.}
\textsc{SWE-Bench} \citep{jimenez2024swebench} and its verified variant supply
$2\,\mathrm{k}+$ GitHub issues but ship every task inside a bespoke Docker
image with all dependencies pre-installed.  
\textsc{DevBench} \citep{li2024devbench} broadens scope to design, coding, and
testing but likewise distributes ready-made containers.  
\textsc{AgentBench} \citep{liu2024agentbench} evaluates multi-step agents across
domains (games, web, reasoning) yet covers only a handful of software tasks and
no system configuration.  SetupBench complements these works by isolating the
\emph{environment bootstrap} phase that precedes any code change.

\paragraph{Environment-setup and DevOps evaluations.}
\textsc{EnvBench} \citep{eliseeva2025envbench} is the closest antecedent,
targeting 994 Python/JVM repos and scoring success via static-analysis or compilation checks. It omits OS-level packages, databases, and
daemon orchestration, which SetupBench tackles explicitly.  
\textsc{LADs} \citep{khan2025lads} proposes an LLM framework for cloud
configuration and bundles a small validation set, while \textsc{OpsEval}
\citep{liu2025opseval} focuses on question-answering in IT operations.
Neither offers runnable, end-to-end setup tasks. SetupBench therefore fills a remaining gap by offering environment-setup instances that span languages, databases, and process managers.

\paragraph{Tool-usage benchmarks.}
\textsc{ToolBench} \citep{qin2024toolllm} and
\textsc{StableToolBench} \citep{guo-etal-2024-stabletoolbench} evaluate an agent’s
ability to call mocked APIs; \textsc{ToolRet} \citep{shi2025retrieval} measures
retrieval of the \emph{right} API.  These datasets abstract away the operating
system entirely.  Earlier CLI-generation work such as \textsc{NL2Bash}
\citep{lin2018nl2bash} focuses on single-line commands.  SetupBench instead
requires multi-command planning, package installation, and service supervision,
bringing tool use closer to real developer workflows. 

In summary, while prior benchmarks illuminate valuable facets of software engineering, none directly evaluate whether an agent can \emph{get the code to run}. SetupBench is designed to fill this evaluation gap.

\section{Conclusion and Future Work}
This paper introduces SetupBench, a comprehensive benchmark evaluating AI agents on real-world software repository setup tasks. We systematically evaluated five leading language models across 93 diverse tasks, providing the first large-scale empirical analysis of agent capabilities in environment-bootstrap tasks—a critical but underexplored aspect of software engineering automation.
Our results reveal both promise and limitations of current coding agents. While the best-performing model (Claude 4) achieved 62.4\% success rate, significant challenges remain. We identified three primary failure modes: failure to install implicit development tooling, hallucinated task constraints leading to unnecessary modifications, and non-persistent environment configurations that break agent-human collaboration workflows. Additionally, all models demonstrated substantial inefficiency with 38-69\% wasted steps due to redundant file reads, poor instruction following, and off-target exploration.

SetupBench establishes environment setup as a distinct and important evaluation domain within software engineering automation. Future work should explore architectural changes including persistent file system representations, semantic search mechanisms, and hybrid approaches balancing efficiency with accuracy, while extending the benchmark to include multi-repository setups and interactive configuration scenarios.

%%%%%%%%%%%%%%%%%%%%%%%%%%%%%%%%%%%%%%%%%%%%%%%%%%%%%%%%%%%%

\bibliographystyle{plainnat}   % or abbrvnat, unsrtnat, etc.
\bibliography{references}      % matches the .bib filename

%%%%%%%%%%%%%%%%%%%%%%%%%%%%%%%%%%%%%%%%%%%%%%%%%%%%%%%%%%%%

\appendix
\section{Example Dataset Entry}

\begin{minted}{json}
{
  "ecosystem": "bundler-compat",
  "base_commit": "45b78114e1a5dab96e59fc70933277a56f65b53b",
  "success_command": "bundle install --jobs=1 --retry=2 --without development test",
  "instance_id": "deps-acts_as_bookable-45b78",
  "problem_statement": "There is a dependency conflict in this project when running bundle install --jobs=1 --retry=2 --without development test. Please resolve the dependency conflict so that the command can run successfully.",
  "task_type": "dependency_resolution",
  "base_image": "ruby:2.7"
}
\end{minted}

\smallskip
\noindent
\textbf{Field conventions.}  
Every entry contains \texttt{instance\_id}, \texttt{problem\_statement},
\texttt{success\_command}, \texttt{base\_image}, and \texttt{task\_type}.  
Category-specific keys include \texttt{ecosystem} (dependency-resolution tasks)
and \texttt{language} (repo-setup tasks).

%-----------------------------------------------------------------------
\section{Repo-Setup Prompt Templates}

\subsection{Setup-Instruction Derivation Prompt}

\begin{minted}{text}
You are tasked with analyzing a GitHub repository and providing detailed
setup instructions for a project. This is part of a benchmark to evaluate
autonomous software engineering agents on their ability to set up projects
correctly. ...

<repo_url> https://github.com/ceph/ceph </repo_url>

1. Thoroughly analyze the repository documentation ...
2. Based on your analysis, provide a detailed, step-by-step guide ...
3. Determine a "success_criteria_command" that can be used to verify ...

<setup_instructions>
[Provide the step-by-step setup instructions here]
</setup_instructions>
<success_criteria_command>
[Provide the single-line command to verify successful setup]
</success_criteria_command>

Note: the sandbox is Ubuntu 22.04 with nothing pre-installed.
\end{minted}

\subsection{Success-Command Synthesis Prompt}

\begin{minted}{text}
You are tasked with creating a success command for a software engineering
benchmark. This command will be used to evaluate whether a repository has been
correctly set up and configured.

Repository URL:
<repo_url>
{{ repo_url }}
</repo_url>

Markdown files from the repository:
<markdown_files>
{{ markdown_files }}
</markdown_files>

Guidelines:
 1. Echo 'Setup successful' or 'Setup failed'.
 2. Only succeed if the repo is fully configured.
 3. Test a key functionality or component.
 4. Chain commands with && and || if necessary.

Provide your answer in the following format:
<success_command>
your bash command here
</success_command>

<explanation>
why this command verifies correct setup
</explanation>
\end{minted}

\medskip
\noindent
All Markdown files in the target repository are injected into the
\texttt{\{\{ markdown\_files \}\}} placeholder, giving the language model rich
context when crafting the validation command.

\section{Dependency-resolution assets}

\subsection{Mining script}

\begin{minted}{python}
#!/usr/bin/env python3
import os
import re
import json
from pathlib import Path
from github import Github

GITHUB_TOKEN = ""
if not GITHUB_TOKEN:
    raise RuntimeError("Please set the GITHUB_TOKEN environment variable")

# Define ecosystems with search queries, error-regex, and lock-files
ECOSYSTEMS = {
    "npm-peer-dep": {
        "search_query":
        "npm ERR! peer dep is:issue in:comments state:closed language:JavaScript",
        "regex": re.compile(r"npm ERR! peer dep", re.IGNORECASE),
        "manifest": "package.json",
        "lockfiles": ["package-lock.json", "yarn.lock"],
    },
    "npm-eresolve": {
        "search_query":
        "npm ERR! code ERESOLVE is:issue in:comments state:closed language:JavaScript",
        "regex": re.compile(r"npm ERR! code ERESOLVE", re.IGNORECASE),
        "manifest": "package.json",
        "lockfiles": ["package-lock.json", "yarn.lock"],
    },
    "pip-conflict": {
        "search_query":
        "ERROR: Could not install is:issue in:comments state:closed language:Python",
        "regex": re.compile(r"ERROR: (?:Could not install|ResolutionImpossible)",
                            re.IGNORECASE),
        "manifest": "requirements.txt",
        "lockfiles": ["Pipfile.lock", "poetry.lock"],
    },
    "poetry-conflict": {
        "search_query":
        "ResolutionImpossible is:issue in:comments state:closed language:Python",
        "regex": re.compile(r"ResolutionImpossible", re.IGNORECASE),
        "manifest": "pyproject.toml",
        "lockfiles": ["poetry.lock"],
    },
    "bundler-compat": {
        "search_query":
        "Bundler could not find compatible versions is:issue in:comments "
        "state:closed language:Ruby",
        "regex": re.compile(r"Bundler could not find compatible versions",
                            re.IGNORECASE),
        "manifest": "Gemfile",
        "lockfiles": ["Gemfile.lock"],
    },
}
\end{minted}
\begin{minted}{python}
MAX_ISSUES = 500
OUTPUT = Path("mined_conflicts.jsonl")

def main():
    gh = Github(GITHUB_TOKEN)
    with OUTPUT.open("a") as out:
        for eco, cfg in ECOSYSTEMS.items():
            print(f"Mining [{eco}]")
            for issue in gh.search_issues(cfg["search_query"],
                                          sort="updated",
                                          order="desc")[:MAX_ISSUES]:
                for comment in issue.get_comments():
                    body = comment.body or ""
                    if not cfg["regex"].search(body):
                        continue

                    # Commit at issue creation
                    default_branch = issue.repository.default_branch
                    commits = issue.repository.get_commits(sha=default_branch,
                    until=issue.created_at)
                    base_commit = commits[0].sha if commits.totalCount else None

                    # Ensure at least one lock-file exists
                    lockfiles_found = []
                    for lf in cfg["lockfiles"]:
                        try:
                            issue.repository.get_contents(lf, ref=base_commit)
                            lockfiles_found.append(lf)
                        except:  # noqa: E722
                            continue
                    if not lockfiles_found:
                        continue

                    snippet = "\n".join(
                        line for line in body.splitlines()
                        if cfg["regex"].search(line)
                    )

                    entry = {
                        "ecosystem": eco,
                        "repo": issue.repository.full_name,
                        "issue_number": issue.number,
                        "issue_url": issue.html_url,
                        "comment_id": comment.id,
                        "snippet": snippet,
                        "matched_at": comment.updated_at.isoformat(),
                        "base_commit": base_commit,
                        "manifest": cfg["manifest"],
                        "lockfiles_found": lockfiles_found
                    }
                    out.write(json.dumps(entry) + "\n")
                    print(f" • Mined {eco} → {entry['repo']}#"
                          f"{entry['issue_number']} @ {base_commit}")

if __name__ == "__main__":
    main()
\end{minted}

\subsection{Validation script}

\begin{minted}{python}
#!/usr/bin/env python3
import os, re, json, tempfile, subprocess, shutil
from pathlib import Path
from concurrent.futures import ThreadPoolExecutor, as_completed
from tqdm import tqdm
from github import Github

ECOSYSTEMS = {
    "npm-eresolve": {
        "image": "node:16",
        "setup": "npm install -g npm@7",
        "cmds": [
            "npm ci --ignore-scripts",
            "npm install --ignore-scripts --legacy-peer-deps"
        ],
        "err": re.compile(r"npm ERR! code ERESOLVE", re.IGNORECASE),
    },
    "npm-peer-dep": {
        "image": "node:16",
        "setup": "npm install -g npm@7",
        "cmds": [
            "npm ci --ignore-scripts",
            "npm install --ignore-scripts --legacy-peer-deps"
        ],
        "err": re.compile(r"npm ERR! peer dep", re.IGNORECASE),
    },
    "pip-conflict": {
        "image": "python:3.9",
        "setup": None,
        "cmds": [
            "pip install --no-build-isolation --no-deps -r requirements.txt",
            "pip install --no-build-isolation -r requirements.txt"
        ],
        "err": re.compile(r"ERROR: (?:Could not install|ResolutionImpossible)",
                          re.IGNORECASE),
    },
    "poetry-conflict": {
        "image": "python:3.9",
        "setup": None,
        "cmds": [
            "pip install poetry && poetry install --no-root "
            "--no-interaction --no-scripts",
            "poetry install --no-root --no-interaction --no-scripts"
        ],
        "err": re.compile(r"ResolutionImpossible", re.IGNORECASE),
    },
    "bundler-compat": {
        "image": "ruby:2.7",
        "setup": None,
        "cmds": [
            "bundle install --jobs=1 --retry=2 --without development test"
        ],
        "err": re.compile(r"Bundler could not find compatible versions",
                          re.IGNORECASE),
    },
}
\end{minted}
\begin{minted}{python}
# CONFIG
GITHUB_TOKEN = ""
INPUT  = Path("mined_conflicts.jsonl")
OUTPUT = Path("validated_results.jsonl")
gh = Github(GITHUB_TOKEN)

def load_entries():
    for line in INPUT.open():
        yield json.loads(line)

def get_done():
    seen = set()
    if OUTPUT.exists():
        for l in OUTPUT.open():
            r = json.loads(l)
            seen.add((r["repo"], r["issue_number"], r["comment_id"]))
    return seen

def record(entry, success, out):
    r = {**entry,
         "validation_success": success,
         "install_output": out.strip(),
         "validated_at": __import__("datetime").datetime.utcnow()
                                                   .isoformat()+"Z"}
    with OUTPUT.open("a") as f:
        f.write(json.dumps(r) + "\n")

def docker_run(workdir, image, cmd):
    full = ["docker", "run", "--rm", "-v", f"{workdir}:/app",
            "-w", "/app", image, "sh", "-c", cmd]
    p = subprocess.run(full, capture_output=True, text=True)
    return p.stdout + p.stderr
\end{minted}
\begin{minted}{python}
def process(entry, done):
    key = (entry["repo"], entry["issue_number"], entry["comment_id"])
    if key in done:
        return

    cfg = ECOSYSTEMS[entry["ecosystem"]]
    tmp = Path(tempfile.mkdtemp(prefix="val_"))
    out, success = "", False

    try:
        repo_url = f"https://github.com/{entry['repo']}.git"
        subprocess.run(["git", "clone", "--depth", "1", repo_url, tmp/"r"],
                       check=True)
        subprocess.run(["git", "-C", tmp/"r", "fetch", "--depth", "1",
                        "origin", entry["base_commit"]], check=True)
        subprocess.run(["git", "-C", tmp/"r", "checkout",
                        entry["base_commit"]], check=True)
        wd = str(tmp/"r")

        if cfg["setup"]:
            out += docker_run(wd, cfg["image"], cfg["setup"])

        for cmd in cfg["cmds"]:
            out += docker_run(wd, cfg["image"], cmd)
            if cfg["err"].search(out):
                success = True
                break
    except Exception as e:
        out += f"\n Pipeline error: {e}"
    finally:
        shutil.rmtree(tmp, ignore_errors=True)

    record(entry, success, out)

if __name__ == "__main__":
    OUTPUT.touch(exist_ok=True)
    done = get_done()
    entries = list(load_entries())
    with ThreadPoolExecutor(max_workers=WORKERS) as ex:
        futures = [ex.submit(process, e, done) for e in entries]
        for _ in tqdm(as_completed(futures),
                      total=len(futures),
                      desc="Validating"):
            pass
    print(" Done; results in", OUTPUT)
\end{minted}
%%%%%%%%%%%%%%%%%%%%%%%%%%%%%%%%%%%%%%%%%%%%%%%%%%%%%%%%%%%%%%%%%%%%%%%%
\setlength{\leftmargini}{1.5em}   % tighter itemize indentation
\section{Database-setup examples: MySQL tier ladder}

\subsection{Tier 1 — basic install + single dump}

\noindent\textbf{Instance ID:} \texttt{dbsetup-mysql-1}

\smallskip
\noindent\textbf{Success command}
\begin{minted}{bash}
mysql -u root -e "USE benchmark_db; SHOW TABLES;" | grep -q products \
  && echo "Setup successful" || echo "Setup failed"
\end{minted}

\noindent\textbf{Task requirements}
\begin{itemize}
  \item Non-interactive MySQL installation with root login.
  \item Create \texttt{benchmark\_db}.
  \item Decompress and import \texttt{dump.sql.gz}.
\end{itemize}

%-----------------------------------------------------------------------
\subsection{Tier 2 — ordered migrations + charset tweak}

\noindent\textbf{Instance ID:} \texttt{dbsetup-mysql-2}

\smallskip
\noindent\textbf{Success command}
\begin{minted}{bash}
mysql -u root -e "USE benchmark_db; SHOW TABLES;" | grep -q products \
  && echo "Setup successful" || echo "Setup failed"
\end{minted}

\noindent\textbf{Task requirements}
\begin{itemize}
  \item Apply numbered \texttt{.sql.gz} migrations with foreign keys.
  \item Ensure server and database use \texttt{utf8mb4}.
  \item Enable root password authentication.
\end{itemize}

%-----------------------------------------------------------------------
\subsection{Tier 3 — port change, strict mode, user permissions}

\noindent\textbf{Instance ID:} \texttt{dbsetup-mysql-3}

\smallskip
\noindent\textbf{Success command}
\begin{minted}{bash}
mysql -u benchmark_user -pbenchmark_pass -e \
  "USE benchmark_db; SELECT COUNT(*) FROM products;" | grep -q '[1-9]' \
  && echo "Setup successful" || echo "Setup failed"
\end{minted}

\noindent\textbf{Task requirements}
\begin{itemize}
  \item Run MySQL on port 3307 (3306 blocked).
  \item Operate under \texttt{STRICT\_TRANS\_TABLES}; patch migrations that
        reference a missing \texttt{DEFINER}.
  \item Re-order and fix out-of-sequence migrations.
  \item Create user \texttt{benchmark\_user/benchmark\_pass} with privileges.
\end{itemize}

%%%%%%%%%%%%%%%%%%%%%%%%%%%%%%%%%%%%%%%%%%%%%%%%%%%%%%%%%%%%%%%%%%%%%%%%
\section{Background-service example: Gunicorn + Unix socket}

\noindent\textbf{Instance ID:} \texttt{bgsetup-gunicorn-systemd-socket}

\smallskip
\noindent\textbf{Success command}
\begin{minted}{bash}
curl --unix-socket /tmp/gunicorn.sock http://localhost/ | grep -q "Hello" \
  && echo "Setup successful" || echo "Setup failed"
\end{minted}

\noindent\textbf{Task requirements}
\begin{itemize}
  \item Install Python, Flask, Gunicorn, and \texttt{supervisord}.
  \item Serve \texttt{/testbed/app.py} via Gunicorn on
        \texttt{/tmp/gunicorn.sock}.
  \item Configure \texttt{supervisord} to restart on failure.
  \item Endpoint must return the string “Hello’’ over the Unix socket.
\end{itemize}

\end{document}